# Electronic Structures of Oxygen-deficient Ta$_2$O$_5$

Yong Yang[1], Ho-Hyun Nahm[1], Osamu Sugino[1,2], and Takahisa Ohno[1]

[1]*National Institute for Materials Science (NIMS), 1-2-1 Sengen, Tsukuba, Ibaraki 305-0047, Japan*

[2]*The Institute for Solid State Physics (ISSP), the University of Tokyo, 5-1-5 Kashiwanoha, Kashiwa, Chiba 277-8581, Japan*

We provide a first-principles description of the crystalline and oxygen-deficient Ta$_2$O$_5$ using refined computational methods and models. By performing calculations on a number of candidate structures, we determined the low-temperature phase and several stable oxygen vacancy configurations, which are notably different from the previous results. The most stable charge-neutral vacancy site induces a shallow level near the bottom of conduction band. Stability of different charge states is studied. Based on the results, we discuss the implications of the level structures on experiments, including the leakage current in Ta$_2$O$_5$-based electronic devices and catalysts.

# I. Introduction

Tantalum pentoxide ($Ta_2O_5$) has attracted considerable attention in recent years due to its potential applications in the electronics industry [1, 2] and catalysis [3, 4]. The high dielectric constant of $Ta_2O_5$ [5-12] puts it a candidate to substitute $SiO_2$ in the conventional complementary metal–oxide–semiconductor (CMOS) devices. The high activity of $Ta_2O_5$ in reducing organic molecules under the UV irradiation makes it a candidate for an effective photocatalyst [3]. Moreover, oxygen-deficient $Ta_2O_5$ has attracted interests as a new non-noble metal cathode material for the polymer electrolyte fuel cell (PEFC) [4]. In these applications, the oxygen vacancy plays a key role: It is considered a source of the leakage current in CMOS and a major reaction center in PEFC. The role of oxygen vacancies in $Ta_2O_5$ is contrasting in CMOS and PEFC applications: The vacancy is *the less the better* for the former while *the more the better* for the latter.

Despite its importance in industrial applications, the basic structural and electronic properties of $Ta_2O_5$ are not well understood. Stephenson and Roth [13] proposed a crystal structure containing 11 formula units (22 Ta and 55 O atoms) for the low-temperature phase (referred to as L-$Ta_2O_5$), where some oxygen sites are partially occupied to satisfy the stoichiometric ratio. It is yet unknown how the oxygen atoms occupy those sites to form the most stable crystalline structure. Later, the structures of simpler crystal models, such as the β-$Ta_2O_5$ [14] and δ-$Ta_2O_5$ [15], were also suggested.

Density functional theory (DFT) calculations were applied to $Ta_2O_5$ on its structural, electronic and dielectric properties within the local density approximation (LDA) or generalized gradient approximation (GGA) [16-19]. The works on oxygen-deficient $Ta_2O_5$ used some simplified crystalline models for the $L$-$Ta_2O_5$ phase to study the energy levels induced by the oxygen vacancies [16, 17]. Recently, both $\beta$-$Ta_2O_5$ [14] and $\delta$-$Ta_2O_5$ [15] were found meta-stable by phonon calculations [18], and the $L$-$Ta_2O_5$ phase was found the most stable. This also cast doubt on the reliability of the simplified models [16, 17]. Moreover, DFT-LDA/GGA severely underestimates the value of band gap [16, 17, 19].

In this work, we revisit the oxygen-deficient $Ta_2O_5$ by providing refined theoretical data on its structural and electronic properties. We overcome the band gap problem encountered in the conventional DFT-LDA/GGA calculations by using hybrid functional for the exchange-correlation interactions, and the instability problem by using a crystal model without simplifying the unit cell of $L$-$Ta_2O_5$ [13]. We have investigated a number of vacancy sites and the associated level structures, which are found distinct from the previous reports.

## II. Computational & Modeling Methods

The calculations are carried out by the Vienna *ab initio* simulation package (VASP) [20, 21], using a plane wave basis set and the projector-augmented-wave (PAW) potentials [22, 23]. The exchange-correlation interactions of electrons are described by two types of functional: The PBE functional [24] and the PBE0 hybrid functional [25]. The energy cutoff for plane waves is 600 eV. For the calculation of unit cell, we use a 4×2×4 k-mesh. For the (2×1×2) supercell (with and without oxygen vacancies), a 1×1×2 k-mesh is employed for structural relaxation and the calculation of electronic density of states (DOS). The k-meshes are generated using the Monkhorst-Pack scheme [26]. The parameters describing the $Ta_2O_5$ unit cell are: $a$ = 6.332 Å, $b$ = 40.921 Å, $c$ = 3.846 Å; $\alpha$ = 90º, $\beta$ = 90º, $\gamma$ = 89.16º. In the experimental L-$Ta_2O_5$ structure model [13], there are two types of partially occupied oxygen sites: four sites with 75% occupation and eight sites with 25% occupation. We have investigated seven plausible occupation patterns by DFT calculations and take the configuration with the lowest total energy as our unit cell model. Generally, the total energy is found lower when the pentagonal bipyramids are separate. More details on constructing the model can be found in the supplementary materials.

The presence of dilute oxygen vacancies is modeled by removing one oxygen atom from the (2×1×2) supercell of $Ta_2O_5$. The vacancy concentration is therefore ~ 0.45% and the corresponding stoichiometric formula for the oxygen-deficient system can be termed as $Ta_2O_{4.9773}$. During structural optimization, both the atomic positions and the supercell geometries are fully relaxed. The formation energy of an oxygen vacancy is calculated by

$$E_{vf} = E\,[Ta_2O_{5-\delta}] + 0.5E\,[O_2] - E\,[Ta_2O_5] - T\Delta S, \qquad (1)$$

where the first three terms are the total energies of the oxygen-deficient $Ta_2O_5$, gas phase $O_2$ (spin triplet state), and the defect-free $Ta_2O_5$, respectively. The value of $\delta$ is 0.0227 in the system under consideration. $\Delta S$ is the entropy change associated with creation of the vacancy. Since the contribution from solid phases is negligible compared to that from gaseous states, $\Delta S$ is approximated by the entropy of the gaseous $O_2$ molecules: We use the value 205.15 $J \cdot K^{-1} \cdot mol^{-1}$ at standard condition [27]. Here, the oxygen vacancy at the charge-neutral, singly and doubly charged states will be investigated. Using the fact that the vacancy has several meta-stable structures that are dependent on the charge state, it is found effective to change the charge state during the geometry optimization to search the stable structure more globally. Therefore, for the charge-neutral systems, we take two procedures to make a comparison. In procedure 1, we relax the structures while keeping the systems charge-neutral. In procedure 2, they are optimized at the doubly positively charged state followed by re-optimization at the charge-neutral state.

## III. Results and Discussion

The building blocks of the L-$Ta_2O_5$ can be classified into two types of polyhedrons centered at Ta [16-18]: Polyhedrons with a pentagonal basal plane (termed as pentagonal bipyramid) and that with a quadrilateral basal plane (termed as octahedron). The polyhedrons are constructed by the Ta and O atoms on the basal planes and two O atoms in between the basal planes. Geometrically, the O atoms can be roughly divided into two groups: the ones on the basal planes (in-plane site) and the ones in between the basal planes (cap sites) [17].

Figure 1 shows the configuration of the oxygen vacancy sites considered in this study. There are two types of the cap-site vacancy: A is the one capping a quadrilateral basal plane and B is the one capping a pentagonal basal plane. Five in-plane sites are considered: C, D and G are the ones coordinated by two quadrilateral basal planes and one pentagonal basal plane; E is the one coordinated by two quadrilateral basal planes, and F is the one coordinated by three quadrilateral basal planes. The formation energies of these oxygen vacancy sites obtained by the two procedures are listed in Table I, where only the charge-neutral state is tabulated because of space limitation. From Table I, one sees that most of the oxygen vacancies at the in-plane sites have lower formation energies and consequently more stable than the ones at the cap sites. The first significant difference from the previous works is that the magnitude of vacancy formation energies is only about one half of the values obtained by the work [17] mentioned above. This indicates that the simplified models in the previous works [16, 17] provide quantitatively different structural properties of the oxygen-deficient system. The configuration D at the in-plane site prepared by procedure 2 turns out to be the most stable. Configurations prepared by procedures 1 and 2 are of almost the same stability except for D and E, where the two configurations prepared by procedure 2 are much more stable than the ones obtained using procedure 1.

The characteristics of the vacancy levels are also displayed in Table I. The cap site vacancies induce shallow states only, in agreement with the previous work [17]. The situation is much different for the in-plane sites vacancies. In the work using a simplified $Ta_2O_5$ model [17], for the most stable vacancy site (in-plane site), the highest occupied vacancy level is deep while it is shallow in our work. As shown below, the deep/shallow characteristics are further confirmed by the more accurate hybrid functional calculations.

To understand the difference resulted from the relaxation procedure 1 and procedure 2, we studied the connection between the total energy of the defective system and the lattice relaxation. The total energy of the relaxed structure can be written as follows

$$E\,[\text{Ta}_2\text{O}_{5\text{-}\delta}] = E\,[\text{Ta}_2\text{O}_{5\text{-}\delta}]_{\text{unrelaxed}} - \Delta E\,, \qquad (2)$$

where $E\,[\text{Ta}_2\text{O}_{5\text{-}\delta}]_{\text{unrelaxed}}$ is the total energy of the corresponding unrelaxed defective structure, and $\Delta E$ is the amount of total energy lowering due to the relaxation of bond lengths and bond angles. The amount of the lattice relaxation is shown in Table II. Here we list the variation of Ta-O bond lengths $R$, the O-Ta-O angles $\alpha$, and the Ta-O-Ta angles $\beta$ referenced to that for an ideal oxygen vacancy (or the crystalline structure), with the parameters $\delta R = \sqrt{\overline{\Delta R^2}} = \sqrt{\frac{1}{N_R}\sum_{i=1}^{N_R}\left(R_i - R_i^0\right)^2}$, $\delta\alpha = \sqrt{\overline{\Delta\alpha^2}} = \sqrt{\frac{1}{N_\alpha}\sum_{j=1}^{N_\alpha}\left(\alpha_j - \alpha_j^0\right)^2}$, $\delta\beta = \sqrt{\overline{\Delta\beta^2}} = \sqrt{\frac{1}{N_\beta}\sum_{k=1}^{N_\beta}(\beta_k - \beta_k^0)^2}$, where the superscript "0" denote the values of the unrelaxed structure. For each vacancy configuration, within the harmonic approximation for the interatomic interactions [28], it should follow that larger values of $\delta R$, $\delta\alpha$, and $\delta\beta$ will lead to lower total energy and consequently lower formation energy and higher stability. It is exactly true for the vacancy configurations B, D, E, and F. For the configurations A, C, and G which have close vacancy formation energies by relaxation procedures 1 and 2, small difference is found for the three parameters. In such cases, the above averaged description of lattice deformation cannot account for the energy difference. More accurate modeling, such as the difference of force constants for the bond length and bond angle distortions at different lattice sites should be considered.

Figure 2(a) shows electronic DOS of the unit cell of L-$Ta_2O_5$. The band gap obtained by the PBE functional is ~ 1.96 eV, which is about ~ 50% of the experimental value (~ 4 eV) [29]. The calculated band gap by the PBE0 hybrid functional is ~ 3.91 eV, comparing quite well with the experimental value. This result points to the role of exchange interactions in determining the excitation levels of electrons in $Ta_2O_5$, since the difference between PBE and PBE0 functional is that 25% of the exchange potential of the latter is replaced by the Hartree-Fock type exact exchange interaction [25].

Figures 2(b) and 2(c) show, respectively, the electronic DOS of the vacancy configuration A, for the structures obtained using the relaxation procedures 1 and 2. For simplicity, the structure obtained by procedure 1 is denoted as A1 and that by procedure 2 as A2. Similar convention applies to the other vacancy configurations. From Figs. 2(b)-2(c), the PBE and PBE0 DOS of A1 and A2 are almost identical. This is consistent with the fact that the formation energy (Table I) of A1 and A2 are almost the same. Although the PBE0 DOS features of both A1 and A2 almost resemble that of the PBE DOS, the most remarkable difference is the band gap (~ 4 eV vs ~ 2 eV). Another significant difference is the appearance of a new vacancy state at ~ 1 eV below the Fermi level in the PBE0 DOS [Fig. 2(c)], which is missing in the PBE DOS [Fig. 2(b)] of both A1 and A2. One of the common features for the PBE and PBE0 DOS is that the highest occupied vacancy level (Fermi level) is located in the conduction band.

The electronic DOS diagrams of the in-plane site vacancy configuration D, the most stable among all the vacancies under consideration, are shown in Fig. 3, for the configuration D1 [Figs. 3(a)-(b)] and D2 [Figs. 3(c)-(d)]. The total DOS obtained by the PBE functional are shown along with that from the PBE0 functional. For the D1 configuration, both the PBE and PBE0 type calculation predict a deep vacancy state which locates near the mid-gap position. The major components of the deep vacancy state are the Ta $5d$ and O $2p$ orbitals; the Ta $6s$ orbital also contributes a minor part [Fig. 3(b)]. Except for the band gap, there is no other notable difference between the PBE and PBE0 DOS of D1. The situation is contrasting in the D2 configuration. Besides the band gap difference, there is newly appearing shallow vacancy state at ~ 0.5 eV below the Fermi level of the PBE0 DOS [Fig. 3(c)], which vanishes in the PBE DOS [see inset of Fig. 3(c)]. Another remarkable difference is that the highest occupied vacancy state is slightly overlapped with the conduction band in D2 while it locates deeply in D1. Again, the major components of the vacancy states come mainly from the Ta $5d$ and O $2p$ orbitals.

As shown above, at charge-neutral state, the Fermi level of vacancy configurations A1, A2, and D2 locates in the conduction band. The DOS peaks of these configurations indicate possible metallic state when the energy dispersion of the vacancy states is wide enough to overlap with the conduction band. When a large number of such oxygen vacancies are introduced, one can expect the onset of metallic conductivity as in the case of metal-insulator transition of the $n$-type degenerate semiconductor [30]. Considering the fact that most oxygen vacancy configurations under consideration (Table I) have the shallow level under the *n*-type condition, one can expect a considerable amount of leakage current when the sample of oxygen-deficient $Ta_2O_5$ is prepared under a nonequilibrium condition. By assuming the coexistence of the various types of vacancies, we may naturally explain the experimental observations that the leakage current in the films of tantalum oxide capacitor persists and depends critically on the local atomic structures around Ta [31].

We then study the stability of $Ta_2O_{5-\delta}$ at different charge states. By neglecting the entropy term [32], the grand potential of the system may be written as

$$\Omega = E[Ta_2O_{5-\delta}]^{\delta q} + (\mu + \delta V)\delta q, \qquad (3)$$

where $E[Ta_2O_{5-\delta}]^{\delta q}$ for $\delta q = 0$ is the total energy of most stable charge-neutral configuration, and for $\delta q = +1$ and $+2$ are the total energies of the relaxed configurations at the corresponding charge states; $\mu$ is the electron chemical potential (i.e., the Fermi level); $\delta V$ is the potential to align the averaged electrostatic potential of the defect system to that of a perfect crystal [33]. In our case we find $0 < \delta V < 0.23$ eV. The calculated grand potential is shown in Fig. 4 for vacancy configurations A and D. For configuration A, when the Fermi level (referenced to the valence top) locates in the interval [3.97 eV, 4.10 eV], the singly charged state is the most stable. Thus, the charge state transition level is $\varepsilon(0/+) = 4.10$ eV and $\varepsilon(+/2+) = 3.97$ eV. There is a small region around the bottom of conduction band, at which the three charge states are nearly of the same stability, with an energy difference of less than 70 meV. For configuration D, The charge state transition level turns out to be $\varepsilon(0/+) = 4.30$ eV and $\varepsilon(+/2+) = 2.76$ eV. Under the *n*-type condition, the oxygen vacancy has two stable charge states $\delta q = 0$ and $+1$, and therefore can accept and donate electrons. Then the electrons required for the electrocatalytic reaction can be supplied via this level. This may explain why the existence of oxygen vacancy enhances the catalytic reaction [34].

The above results may explain the experimental finding that the Fermi level of the oxygen-deficient $Ta_2O_5$ locates near the bottom of the conduction band (i.e., the singly charged state of A and D) [35]. The observation that the vacancy created by Ar ion bombardment induces a vacancy band at around the mid-gap position [36] may correspond to the vacancy configurations D1 or F1or F2. For the singly charged state of A and D, despite the presence of one unpaired electron in each supercell, spin-polarized calculations show zero magnetic moment on each atom.

To summarize, we have investigated the electronic structures of oxygen-deficient $Ta_2O_5$ using DFT calculations. Based on refined computational methods and models, the electronic properties of the defective system are found to be notably different from previous studies. The cap site vacancy induces shallow electronic states while the in-plane site vacancy can induce either shallow or deep states. The most stable charge-neutral configuration shows a shallow vacancy state. The transition between different charge states is studied. The results are used to explain available experimental findings.

**Acknowledgments:** This work is supported by the Global Research Center for Environment and Energy based on Nanomaterials Science (GREEN) at National

Institute for Materials Science (NIMS). The first-principles calculations were carried out by the supercomputer (SGI Altix) of NIMS.

Table I. Vacancy formation energy of different types of oxygen vacancy in $Ta_2O_5$. The formation energies for configurations optimized using procedure 1 and procedure 2 are labeled by $E_{vf1}$ and $E_{vf2}$, respectively. The characteristics of vacancy level (Vac Level) for each configuration are also listed.

| Vacancy Configurations | Cap Site | | In-plane Site | | | | |
|---|---|---|---|---|---|---|---|
| | A | B | C | D | E | F | G |
| $E_{vf1}$ (eV) | 4.907 | 5.389 | 4.301 | 4.164 | 4.702 | 5.185 | 3.878 |
| Vac Level | shallow | shallow | deep | deep | shallow | deep | shallow |
| $E_{vf2}$ (eV) | 4.914 | 5.351 | 4.268 | 3.495 | 3.657 | 5.026 | 3.898 |
| Vac Level | shallow | shallow | shallow | shallow | shallow | deep | shallow |

Table II. Quantities characterizing the lattice relaxation of the vacancy configurations with reference to the unrelaxed $Ta_2O_{5-\delta}$ structure.

| Vacancy Configurations | | Cap Site | | In-plane Site | | | | |
|---|---|---|---|---|---|---|---|---|
| | | A | B | C | D | E | F | G |
| Relaxed by Procedure 1 | δR (Å) | 0.038 | 0.025 | 0.041 | 0.035 | 0.043 | 0.028 | 0.059 |
| | δα (º) | 2.282 | 1.368 | 2.502 | 2.091 | 2.741 | 2.033 | 3.898 |
| | δβ (º) | 3.304 | 1.799 | 3.619 | 1.992 | 2.565 | 2.481 | 4.349 |
| Relaxed by Procedure 2 | δR (Å) | 0.040 | 0.033 | 0.045 | 0.066 | 0.102 | 0.040 | 0.060 |
| | δα (º) | 2.359 | 1.925 | 3.226 | 4.865 | 5.760 | 2.959 | 3.968 |
| | δβ (º) | 3.389 | 2.313 | 3.374 | 6.905 | 6.972 | 3.699 | 4.463 |

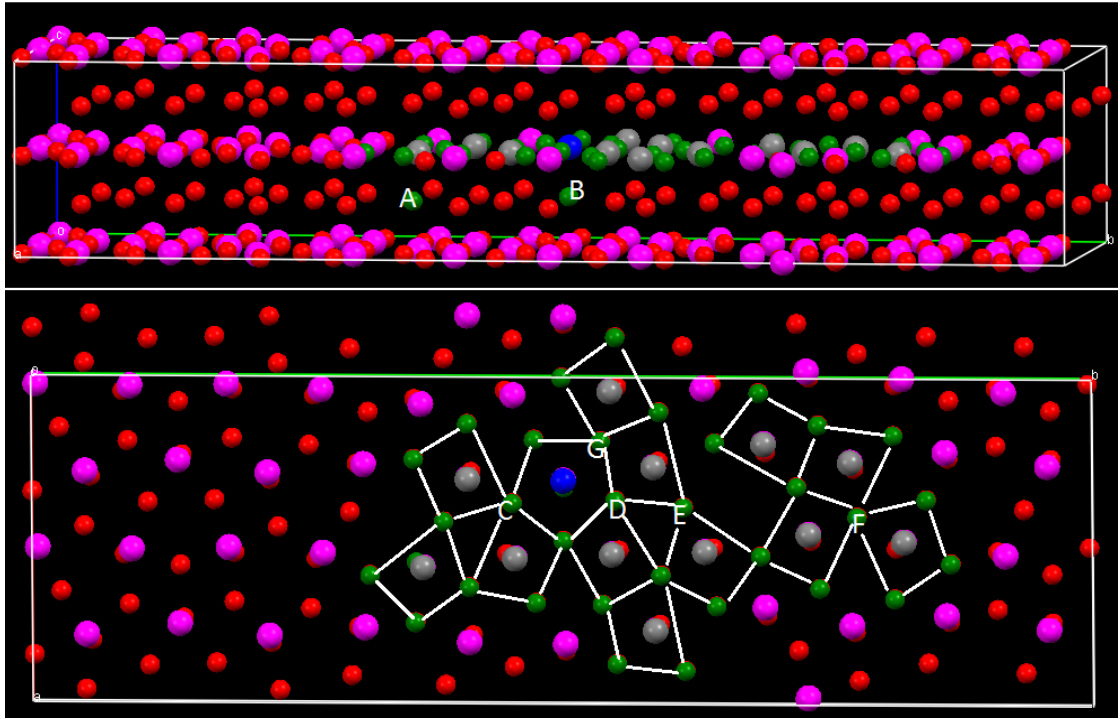

**FIG. 1 (Color online)** Schematic diagram for the cap site oxygen vacancy configurations A, B (upper panel), and the in-plane site vacancy configurations C, D, E, F, G (lower panel) in a (2×1×2) supercell of $Ta_2O_5$. The quadrilateral and pentagonal basal planes are marked. The O atoms are represented by smaller spheres (green for the O atoms of polygon basal planes near the vacancy sites and red for the others) and the Ta atoms are represented by larger spheres (gray for the quadrilateral centers and blue for the pentagonal centers near the vacancy sites, and pink for the others). The Ta-O bonding is considered effective when the Ta-O distance is no more than 2.5 Å. The number of quadrilateral and pentagonal and planes will change slightly when the effective bonding length varies, e.g., at the Ta-O length $\leq 2.6$ Å, the quadrilateral basal plane shared by vacancy sites D, E and G will switch to a pentagonal one.

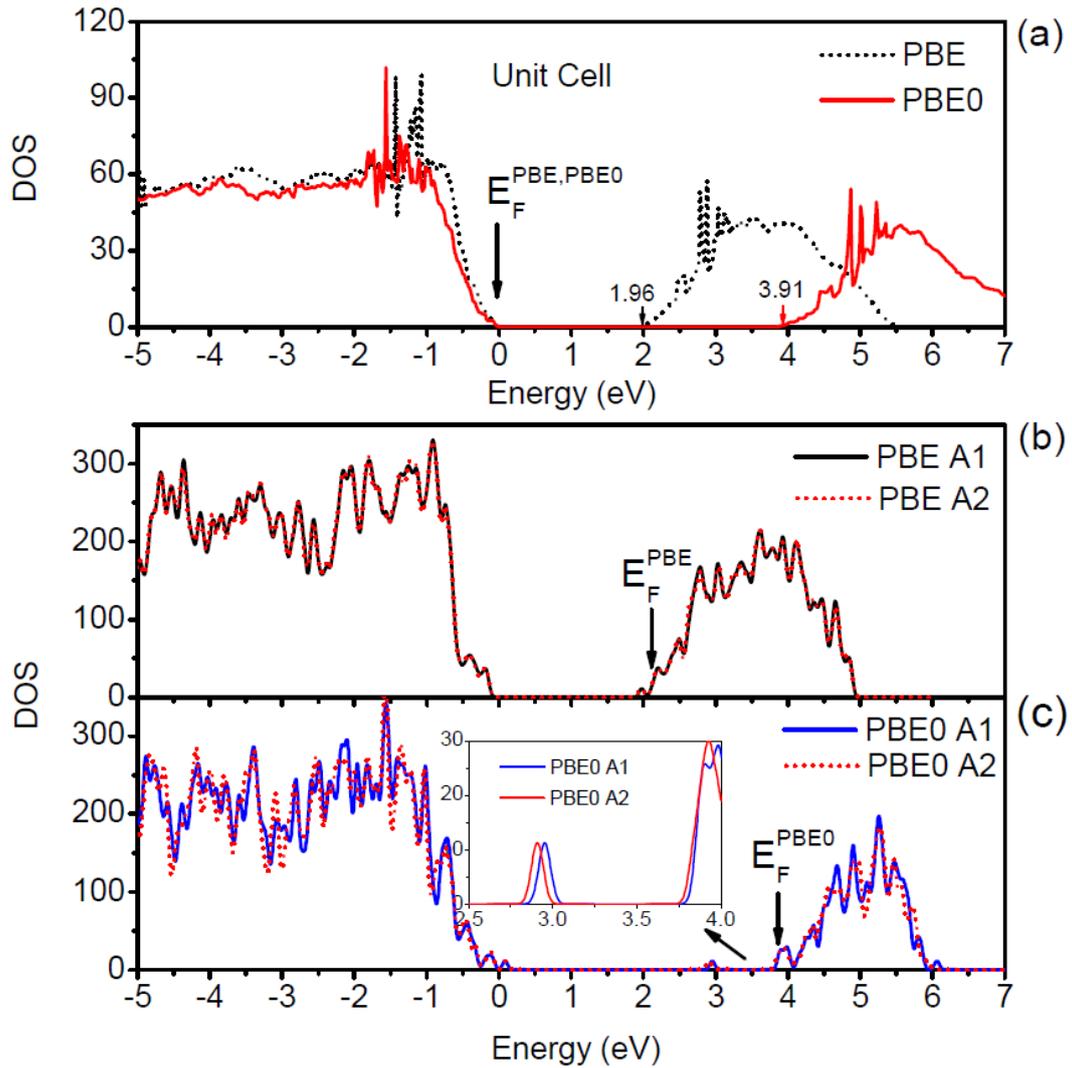

**FIG. 2 (Color online) (a)** Calculated electronic density of states (DOS) of L-Ta$_2$O$_5$ unit cell, by using the PBE and the PBE0 functional. **(b)** Calculated DOS by PBE functional for the configurations A1 and A2. **(c)** Calculated DOS by PBE0 functional for the configurations A1 and A2. The DOS peaks near the bottom of conduction band are highlighted in the inset. The unit for DOS is states/eV/supercell. The top of valence band is set at 0. The position of the Fermi level (E$_F$) is marked by a vertical arrow. The convention applies to all the following figures.

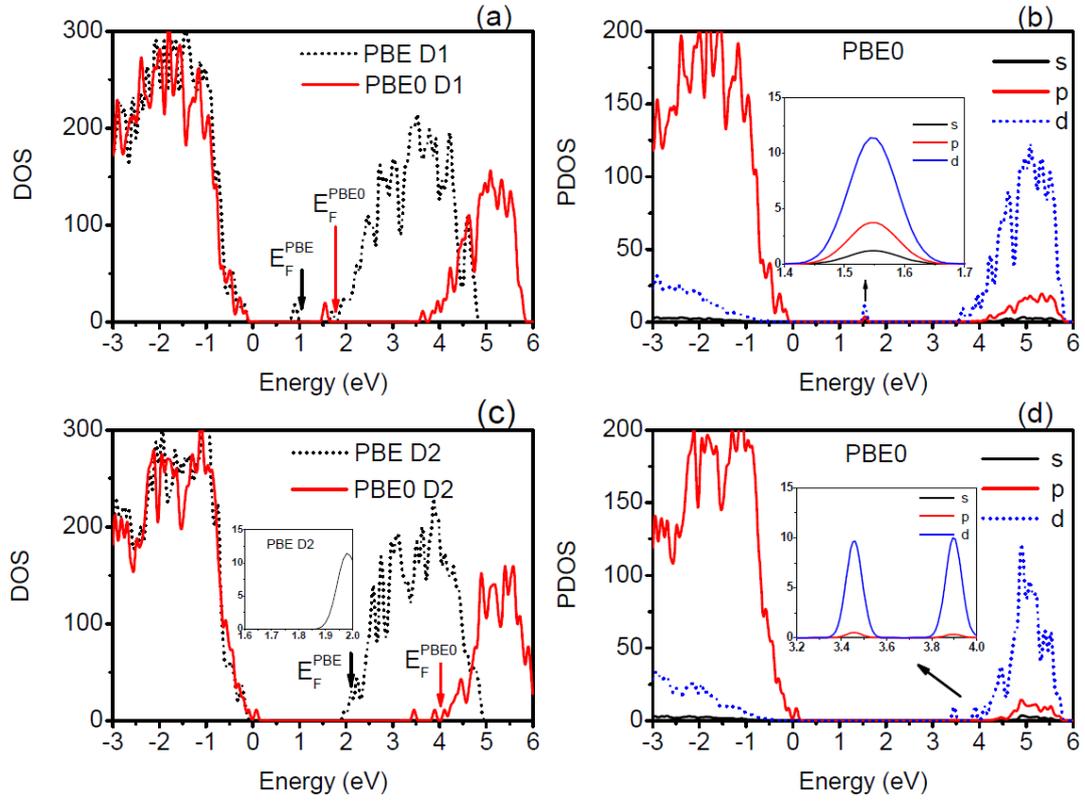

**FIG. 3 (Color online)** Calculated total (left panels) and partial (right panels) electronic DOS of the vacancy configurations D1 (upper panels) and D2 (lower panels) by using the PBE and PBE0 functional.

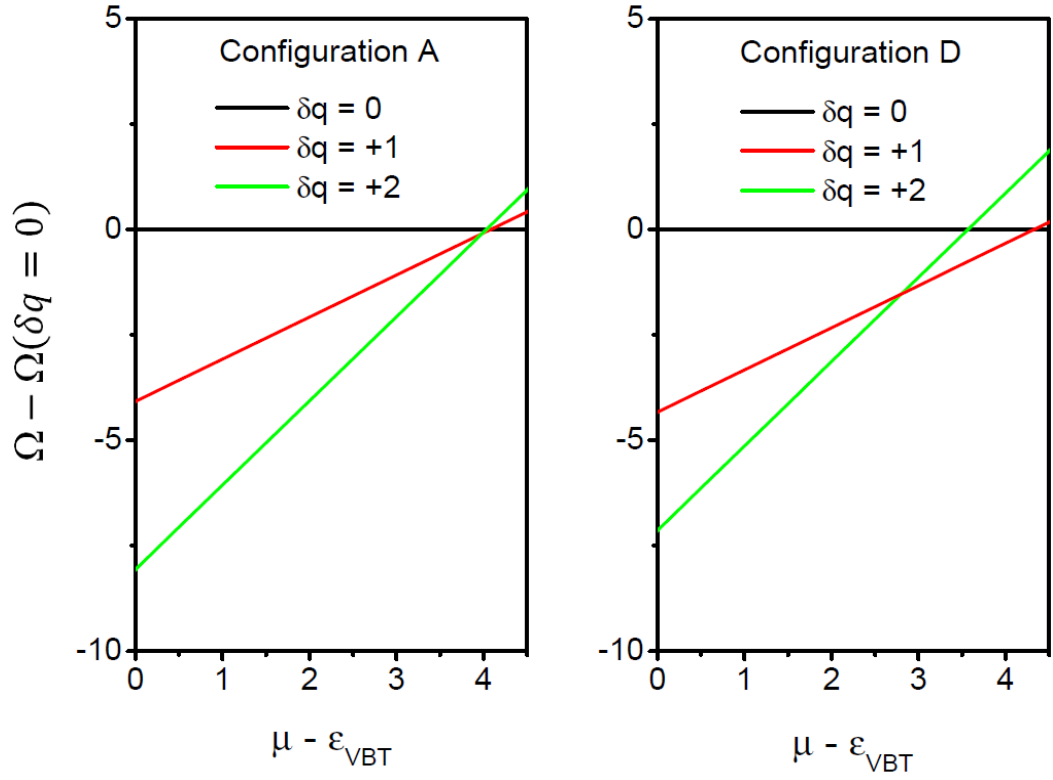

**FIG. 4 (Color online)** Calculated grand potentials for the charge states of vacancy configurations A and D of the $Ta_2O_{5-\delta}$ system: $\delta q = 0$, +1 and +2. The unit for energy is eV. The symbol $\varepsilon_{VBT}$ is the top of valence band.

**Supplementary Materials**

## Construction of the Crystal Model for Ta$_2$O$_5$

Y. Yang, H.-H. Nahm, O. Sugino, and T. Ohno

We directly use the orthorhombic tantalum pentoxide (Ta$_2$O$_5$) with p2mm space group (formula units Z=11) which has two types of partially occupied oxygen (O) sites: four O atoms (0.75) and eight O atoms (0.25) (Fig. S1). That is, three of four O atoms (0.75) and two of eight O atoms should be selected.

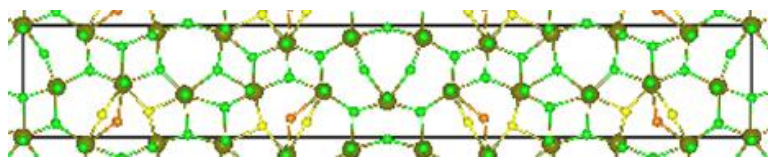

**FIG. S1** Experimentally identified p2mm lattice structures (Z=11). Gold circle: Ta; green circle: oxygen (1); orange circle: oxygen (0.75) (one of four O atoms should be excluded); yellow circle: oxygen (0.25) (two of eight O atoms should be included).

Here, we systematically tested all of possible combinations, and when three of O atoms (0.75) are selected, each 0.25 O atom, which is much close to each 0.75 O atom, is excluded because the case is energetically metastable (see Fig. S2). We note that above selection cannot conserve the symmetry of p2mm-Ta$_2$O$_5$ with 4 symmetric operations (resulting in the breaking of mirror symmetry along a long axis), because the portions of partial occupancy sites cannot be selected with the symmetry conservation.

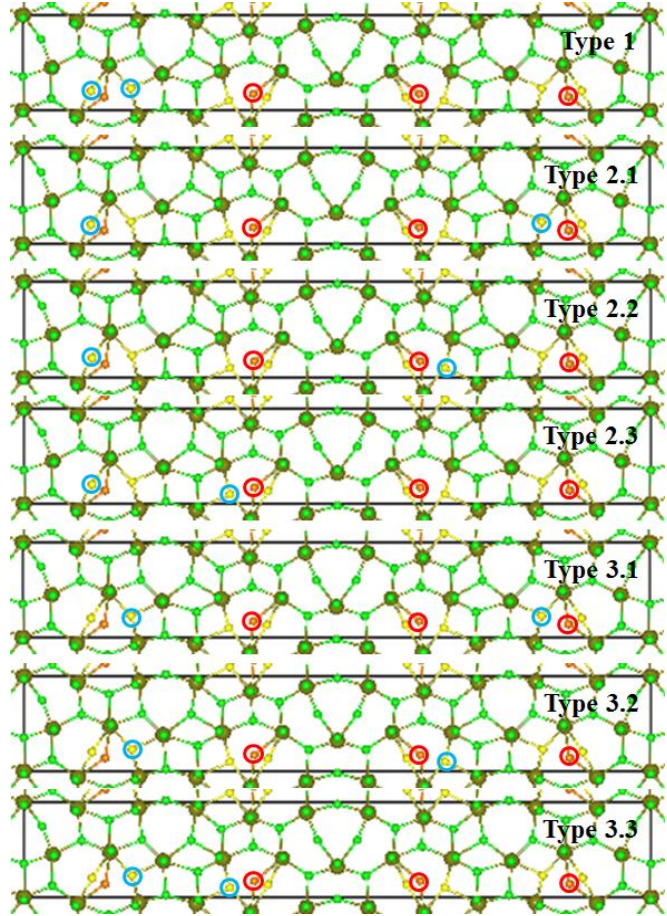

**FIG. S2** Possible configurations for the partially occupied O atoms.

For above seven cases, we compared the energetics of the final structures, as listed in Table SI. It is found that except Type 2.3 and Type 3.2, all of remnant five cases are stable because the stable types have a similar stacking pattern as well as the characteristic polyhedrons, consisting of 20 octahedrons and 2 pentagonal bipyramids (Fig. S3). So, we select Type 1 as a representative of p2mm-$Ta_2O_5$.

**Table SI** Total energies of the $Ta_2O_5$ unit cells under consideration.

| Unit cell | Type 1 | Type 2.1 | Type 2.2 | Type 2.3 | Type 3.1 | Type 3.2 | Type 3.3 |
|---|---|---|---|---|---|---|---|
| $E_{tot}$ (eV/cell) | -749.366 | -749.365 | -749.311 | -745.145 | -749.361 | -748.307 | -749.292 |
| $E_{tot}$(eV/$Ta_2O_5$) | -68.124 | -68.124 | -68.119 | -67.740 | -68.124 | -68.028 | -68.117 |

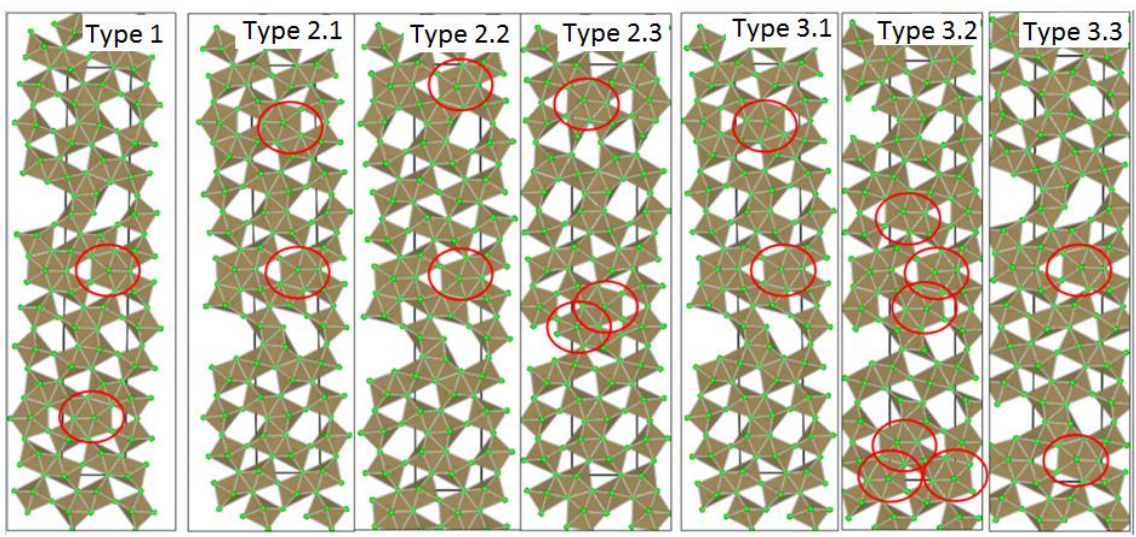

**FIG. S3** Final structures of the seven configurations.

X-ray Diffraction (XRD): In order to compare the experimental XRD data with present identified crystal structure, we calculated the XRD by using a RETAIN-FP program included in VESTA program (see Fig. S4)

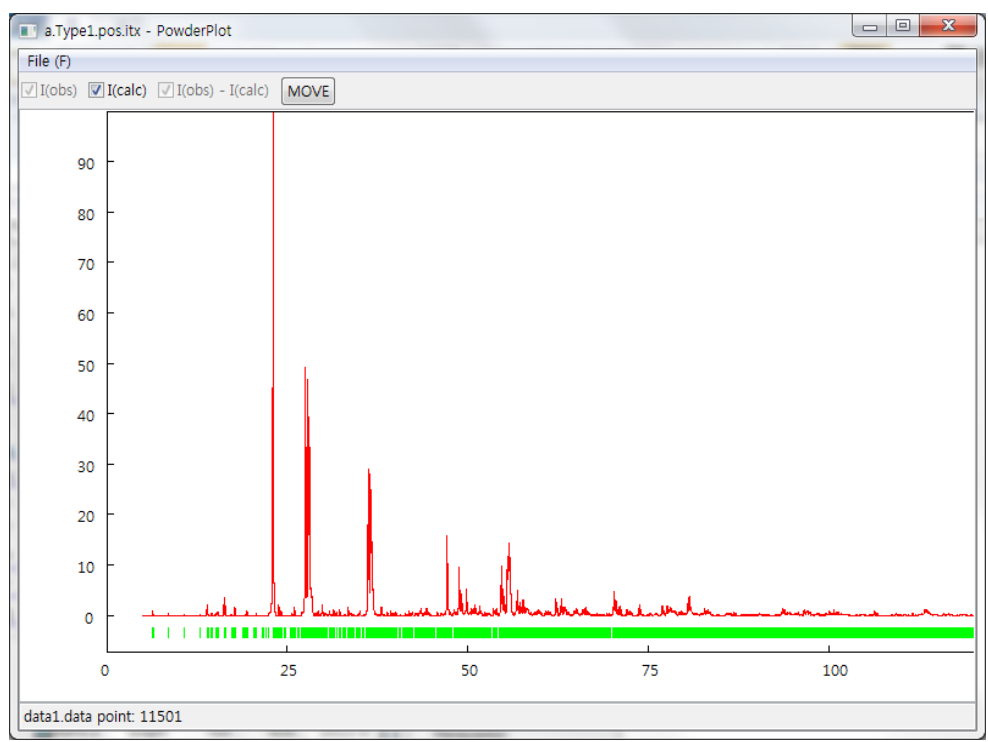

**FIG. S4** The XRD data of Type 1 $Ta_2O_5$.